# Risk Assessment, Threat Modeling and Security Testing in SDLC

1st Alya Hannah Ahmad Kamal
*School of Computer Science & Engineering,*
*Taylor's University*
Selangor, Malaysia
alyahannah@gmail.com

2nd Caryn Chuah Yi Yen
*School of Computer Science & Engineering,*
*Taylor's University*
Selangor, Malaysia
carynchuah3@gmail.com

3rd Gan Jia Hui
*School of Computer Science & Engineering,*
*Taylor's University*
Selangor, Malaysia
jiahui662@gmail.com

4th Pang Sze Ling
*School of Computer Science & Engineering,*
*Taylor's University*
Selangor, Malaysia
szelingpang46@gmail.com

5th Fatima-tuz-Zahra
*School of Computer Science & Engineering,*
*Taylor's University*
Selangor, Malaysia
fatemah.tuz.zahra@gmail.com

**Abstract** – The software development process is considered as one of the key guidelines in the creation of said software and this approach is necessary for providing a more efficient yet satisfactory output. Without separation of work into distinct stages, it may lead to many delays and inefficiency of the project process where this disorganization can directly affect the product quality and reliability. Moreover, with this methodology established as the standard for any project, there are bound to be missteps specifically in regard to the involvement of security due to the lack of awareness. Therefore, the aim of this research is to identify and elaborate the findings and understanding of the security integrated into the process of software development as well as the related individual roles in ensuring that this security is maintained. Through thorough analysis and review of literature, an effort has been made through this paper to showcase the correct processes and ways for securing the software development process. At the same time, certain issues that pertain to this subject have been discussed together with proposing appropriate solutions. Furthermore, in depth discussion is carried out regarding methods such as security testing, risk assessment, threat modeling and other techniques that are able to create a more secure environment and systematic approach in a software development process.

## 1 Introduction

Commercial applications and websites are not totally safe, there is always the existence of numerous flaws and bugs that can be exploited by attackers [1][2] and lead to financial loss [3] along with security and privacy invasion. This is not a small problem. Every month, there are at least 2 reports of cyber-attacks to websites or databases around the world, millions of dollars are lost either directly or indirectly. Until the latest vulnerability found and exploited, vendors only then provide patches for that particular vulnerability. However, the loss has been caused. Making matters worse, the patches may open to new possibilities of attacks. Also, system administrators will have a hard time keeping up with the latest patches. Prevention is better than cure. Normally, developers will be examining the final production code to look for possible vulnerabilities and bugs, then fix it before releasing into the market. However, there is a possibility where the final product consists of too many problems that require high skills and high costs to fix. Hence, the best solution would be integrating security principles into the product from the start of the software development process. A methodology is proposed where security principles should be applied at every development stage and each stage will be tested for compliance with those security principles. This methodology is well-known as the secure software development process, resulting in a much safer end product.

The secure software development process is also known as the Secure Software Development Life Cycle (SDLC) [4]. SDLC is a framework for the process used by organizations in building an application from scratch until the end. In general, SDLCs consist of six phases, which are requirements and planning, design, implementation, testing, deployment and maintenance. The six phases will be discussed in-depth in section 2. It has a structured flow of stages that assist organizations to produce high-quality software products efficiently [5]. It provides a fine-grain analysis of each stage of the process to ensure quality. SDLC is a plan that starts with an existing system evaluation to find any deficiencies. Next, the requirements of the new system are defined. After defining, the creation of the software product goes through designing, developing/ implementing, testing and deployment. Finally, the product is maintained. During the whole process, developers will be getting feedback from voluntary end-users and stakeholders to eliminate redundant rework. SDLC is a repetitive methodology, developers must ensure the quality of the code at every cycle. Popular SDLC models include the waterfall model, spiral model and agile model. Most organizations tend to put less focus on testing, but little do they know that a stronger focus on testing is able to save lots of rework, time and cost.

Secure SDLC (Fig. 1) refers to integrating security checks as early as the first phase of the SDLC [6]. Developers need to be aware of potential security concerns at every stage of the SDLC to ensure the security of the software product. The security principles that will be integrated into each phase will be further elaborated in section 3.1. An example of an established security lifecycle is Microsoft SDL, which is a spiral model of SDLC proposed by Microsoft. It has two additional extensions compared to the common SDLC model with six stages. This model has seven stages, including training, requirements, design, implementation, verification, release and response.

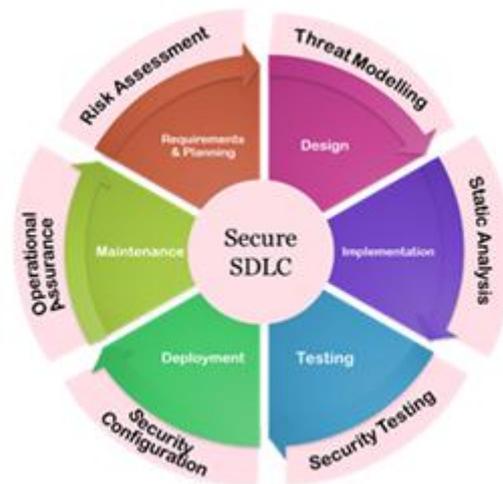

Fig. 1. Secure software development life cycle

Secure SDLC refers to integrating security checks as early as the first phase of the SDLC [6]. Developers need to be aware of potential security concerns at every stage of the SDLC to ensure the security of the software product. The security principles that will be integrated into each phase will be further elaborated in section 3.1. An example of an established security lifecycle is Microsoft SDL, which is a spiral model of SDLC proposed by Microsoft. It has two additional extensions compared to the common SDLC model with six stages. This model has seven stages, including training, requirements, design, implementation, verification, release and response.

Secure SDLC is important because it enables the team to identify security issues at every development phase, not only during the last stage. By implementing secure SDLC, the product will be improved and more secure because developers are fixing found security vulnerabilities at every stage. Security vulnerabilities of the final product will be harder to find, and even if it is found, the impact of it is minimized [6]. Secure SDLC's purpose is to not eliminate traditional checks like penetration tests. It is to include the security concept across the whole process, not just at the end. This leads to vulnerabilities being discovered and reduced early by effectively building security in. Having a strong and secure SDLC is essential to ensure the final product is not subject to attacks by hackers and any other malicious users. If not done so, it can lead to increased deployments of insecure software in sensitive domains like healthcare, unmanned vehicles [7] and mission critical applications.

This paper provides a comprehensive survey of the secure software development process for developers. This section has briefly introduced secure software development lifecycle. Section 2 discusses the literature review which includes the phases of SDLC and the security principles integrated into it. Also, it further elaborates the security principles such as secure code review, risk assessment, threat modeling and security testing. Section 3 provides the methodology used to collect results. In section 4 an in-depth discussion on unique findings is carried out and the solutions that can be used to solve the issues and challenges faced by developers are discussed. Section 5 concludes the paper with summarized findings.

## 2  Literature Review

As mentioned in section 1, it is important to understand the significance of the security aspect in software development process. This is inclusive of the Software Development Life Cycle (SDLC) itself, secure code review, risk assessment, security testing and threat modeling. Comprehensive guidelines should be provided to the developers to ensure that they are updated on the latest information on the process to a secure software system. In this section, the security aspect and steps required for integrating them in various development phases are thoroughly discussed. Before that, it is necessary to understand the roles of each phase of the SDLC in more detail. There are six phases in SDLC, which are: 1) Requirements and Planning, 2) Design, 3) Implementation, 4) Testing, 5) Deployment and 6) Maintenance, respectively. Each phase has its own security that should be implemented.

Requirements and Planning phase is where the project team receives all project details from the stakeholders or client. All requirements needed in the software development project will be defined [8]. In this phase, the development team has to establish security requirements and does a security risk assessment [9]. This helps in identifying what security is needed for the software. Moving on to the Design phase, this phase includes all activities that will occur prior to coding. The design approach consisting of architectural modules of the software and data flow diagram will be proposed according to the initial plan and requirements [10]. The project team has to apply secure design patterns on design and architecture. Besides, the team should analyze the attack surface of the software in order to have an overall idea of the software and is able to identify the security vulnerabilities. Threat modeling is another activity that needs to be implemented in the Design phase. It helps in detecting potential threats and studying risks before producing a mitigation plan.

The next phase of SDLC is the Implementation phase. The team has to decide on the coding standards and style such as the naming of variables and methods [11]. With this, a consistent code will be produced and will be more understandable to members who are in charge of the code testing. During the coding of the software, the team needs to be aware of buffer overflows as this has been one of the most common attacks in the industry [9]. Secure code review and static analysis should be integrated and these two security measures will be discussed after. In the Testing phase, programmers examine the source code for deficiencies and errors. Bugs fixing and retesting will continue until the software achieves the security requirements and specifications. It is important to apply dynamic analysis and penetration testing in this phase as it will help in identifying security issues. Now that the code review and testing are done, the software is ready to be deployed. The software will be released into the market and security measures such as networking and cryptography and permissions will be integrated during this Deployment phase. Lastly, we have the Maintenance phase. After receiving feedback from users, maintenance will be done frequently to improve user experience. Patching, regression and assessment will take place while the team updates the software. An incident response plan should also be planned beforehand so that it can be executed immediately when the software is attacked. SDLC is applicable in various environments as well as modern development approaches, such as distributed software development [12] environment and component-based development [13] which can lead to insecure products if not monitored and built carefully. Therefore, it is necessary to integrate security features in each phase of the software development process.

### 2.1  Risk assessment

Risk is the likelihood of a threat-source to exploit vulnerabilities, which results in the impact to the target [14]. In order to mitigate risk in a product or service, risk management is integrated into the SDLC. The two major activities of risk management are risk assessment and risk reporting [15]. In this paper, we will be focusing on risk assessment, which is the first step in risk management, that is integrated in the planning and requirements phase. The identification of risk, analysis of risk, and prioritizing risk are the main activities of risk assessment.

Risk assessment is used to determine the level of the potential threat and the risk integrated in the IT system throughout the SDLC [14]. Risk identified in the risk assessment process can be reduced or eliminated by applying suitable controls during the risk mitigation process. Usually, the risk assessment is followed by threat modeling which will be explained further in section 2.2.

In the risk assessing process, the first step is to assume the software will be attacked and think of the factors that motivates the threat actor [16]. List out the factors such as the data value of the program, the security level of companies who provide resources of the code depends on, the clients purchasing the software, and how big is the software distributed (single, small workgroup or released worldwide). Based on the factors, write down the level of risk that is acceptable. For instance, a data loss may cause the company to lose $2000, but to eliminate all potential security bugs in the code may cost $20,000. The company and some other panels have to decide whether it is worth it. On the other hand, if the attack causes damage to the company's reputation which costs the company more in the long run compared to fixing the code, then it may be worth the fix.

The next step is risk evaluating. Include factors into considerations, such as the worst-case scenario if the attacker has successfully attacked the software. Determine the value of data to be stolen, valuable data such as user's identity, credentials to gain control of the computer, or just useless data. Another factor to consider is the difficulty to mount a successful attack. The level of risk is acceptable if the attacker does small harm, such as installing a Trojan horse on the computer that can take advantage of the lucky draw of a company's annual dinner. On the other hand, the level of risk that is high will not be acceptable, and best to mitigate. For example, a vulnerability that can be exploited by anyone running prewritten attack scripts or using botnets to spread the scripts to compromise computers and networks. The number of users that will be affected is a factor as well. Some attacks only affect one or two users, but the denial of service attack will affect thousands of users when a server is attacked. Moreover, thousands of computers may be infected by the spread of worms. Last factor is the accessibility of the target. Determine that the target accepts requests across a network or only local access, authentication is needed for connection establishment or anyone can send requests.

Risk assessment is to identify the likeliness of the software being attacked and how big is the damage. After identifying, companies need to find out how the attack will be performed and the target to be attacked. This can be done by creating a threat model and applying threat modeling methodologies, which is further elaborated in section 2.2.

## 2.2 Threat modelling

Threat modeling is best integrated in the design phase of an SDLC, before any code is written. Threat modeling is a structured process of identifying potential security threats, and to prioritize techniques to mitigate attacks so that something valuable such as confidential data is protected [17]. Threat modeling lets security teams to have an analysis of the security controls needed based on information acquired, such as the current information systems and the threat landscape which consists of possible attacks, the methodology and the target system. Threat modeling is the key to a focused defense. It brings great advantage when performed at early stages, potential issues can be found early and solved which saves fixing costs down the line. It reduces threats from the start of the process [18].

Over the years, more threat-modeling methods are being developed. Not all the methods have the same purpose, some focus on risk or privacy concerns while some are people centric [18]. These methods can be combined to have a better view of potential threats. While adopting a threat methodology or framework, it is important for the professionals or security team to analyze which method aligns with their business goals [17]. The three most common threat modeling methodologies are STRIDE, DREAD, and PASTA.

STRIDE is a widely used threat model developed by Microsoft which evaluates the detailed design of a system. STRIDE can be used to identify the types of threats as it covers the six main threats, which is shown in the table above, including the property violated by the threat and the threat definition. In this model, the system's data flow diagram is to be developed and each node is applied with the STRIDE model [19]. The identification of security threats is a manual process which is not supported by tools. Using data flow diagrams and integrating STRIDE, the system entities, events and known boundaries can be identified [18]. The threats which are a part of STRIDE model are described in Table 1 along with the threat domains. The DREAD model can be applied after identifying the threats.

Table 1. Threats, property violations and threat definitions included in STRIDE.

|   | Threat | Property Violated | Threat Definition |
|---|---|---|---|
| S | Spoofing identity | Authentication | Pretending to be something or someone other than oneself |
| T | Tampering with data | Integrity | Modifying something on disk, network, memory or elsewhere |
| R | Repudiation | Non-repudiation | Not claiming responsibility of an action: could be false or true |
| I | Information disclosure | Confidentiality | Providing information to an unauthorized body |
| D | Denial of service | Availability | Exhausting resources which are need for service provision |
| E | Elevation of privilege | Authorization | Allowing someone to do something they are not authorized to do |

DREAD is also a methodology created by Microsoft which can be an add-on to the STRIDE model [20]. DREAD is a model that ranks threats, by assigning identified threats according to the severity and priority level [19]. Without STRIDE, the DREAD model also can be used in assessing, analyzing and finding the risk probability by threat rating. The abbreviation DREAD stands for five questions about each potential (shown in Table 2) threat which is covered in the table above. In the process, ratings between one and three are used to rate the questions [20].

Table 2. Brief description of components in DREAD methodology.

| Name | Description |
|---|---|
| **D**amage potential | How much damage can this threat do to the system? |
| **R**eproducibility | Is it easy to reproduce? |
| **E**xploitability | How much effort and experience does it require to be exploited? |
| **A**ffected users | In case the threat leads to an attack, how many users are potentially going to be affected? |
| **D**iscoverability | How easy is it to discover the vulnerability? |

PASTA is short for Process for Attack Simulation and Threat Analysis, a threat modeling framework that is risk centric [18]. PASTA has seven stages as shown in the Fig. 2. Its focus is to align technical requirements with business objectives. During different stages, different elicitation tools and designs are used. PASTA involves the threat modeling process from analyzing threats to finding ways to mitigate them, but on a more strategic level. It identifies the threat, enumerates the threats, then assigns them a score. This helps organizations to find suitable countermeasures to be deployed in order to mitigate security threats [17].

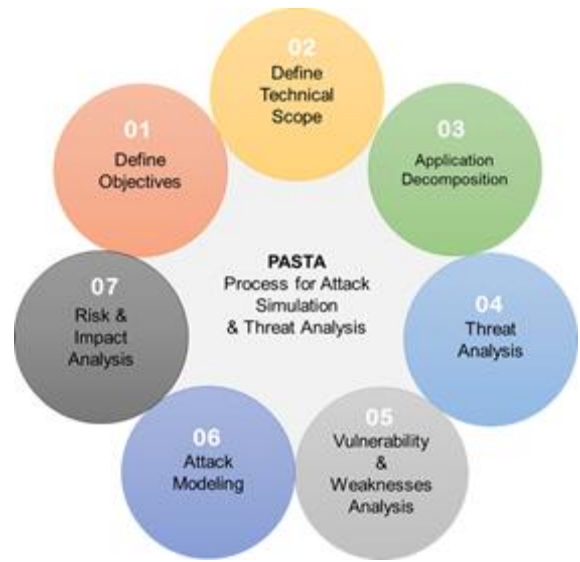

Fig. 2. Process for attack simulation and threat analysis

### 2.3 Secure code review

With the implementation of a secure code review in the phases of an SDLC especially during the implementation phase, it will surely increase the robustness of the product without bearing any additional cost for future patches. Secure code review is defined as a measure where the code itself is verified and validated to ensure vulnerabilities that are found can be swiftly mitigated and removed to avoid creating an insecure software [21]. This implementation of having the developers be aware and proactive in reviewing the codes during development can result in faster mitigation response as well as fewer threats or vulnerabilities to be left unattended [22]. Reviewing code is a crucial step in the SDLC as developers then overlook this approach and mainly focus on only the functionality aspect of the production of the software. With that in mind, there are two paths in which a code can be reviewed for security assurance, be it the black box or white box approach.

The white box technique is a manual inspection of the code by the developers themselves or a security analyst [21]. This manual-based approach is further explained and can be found in subsection 2.4 of this topic. Moving on to the black box technique, this technique is with the utilization of automated tools in identifying the vulnerabilities in a code during the development phase. This technique can be very useful to the related parties especially when analyzing large sets of code, with this type of tool, they are able to identify the possible code insecurities swiftly compared to when it is done manually. Even so, a combination of both techniques can be done and may prove most beneficial depending on the project and code magnitude.

With whichever approach applied, it will promptly assist in covering vulnerabilities and threats that can be found affecting areas such as the logic flow, error handling, validation, session management, the confidentiality, integrity and availability of the future product, the known vulnerabilities and many others [23]. By applying one factor during the development process can in turn protect the product from these threats without getting any significant loss from doing so. Not only that, this will assist in uncovering possible errors in the design or framework of the system that are unable to be identified by just tools as mentioned previously, and also reduce the number of false positives that may occur.

Of course, with the mitigation of the vulnerabilities and threats that could have proven troublesome is the main perk of implementing a secure code review, there are other many positive outcomes besides that. The other benefits include the work effort that can be significantly reduced yet quality being maintained due to the earlier stage being the space where the errors found is mitigated. This can bring about less complexity and tedious work needed to be done if the error was instead identified after the product is in the production and release phase. It ultimately prevents any setbacks on the release and issues that are exposed to the users. Other than that, the cost of the project itself will not be disrupted or increased as the effort put in is proportional to the cost put into this project. By implementing this approach, the organization will also be compliant to the standards that were set and lastly, as mentioned earlier on in this section, the organization's reputation and reliability will increase in the industry which can be greatly beneficial [21]. All in all, by doing this, the developer can not only be aware of vulnerabilities and insecurities found quicker but also be able to increase the overall visibility of the code in the application layer and be cost effective, efficient as well as properly mitigating the issues early for a more systematic and smooth development process.

## 2.4 Security testing

Security testing plays the main role in achieving security in SDLC and should be implemented in all phases. According to Michael Felderer and Ruth Breu, security testing does the validation of security requirements related to confidentiality, integrity, availability, authentication, authorization and non-repudiation [24]. Security testing assesses a system, software or web application for vulnerabilities and other attack vectors. It is composed of many processes, making sure that the code of the application is functioning well as planned and does not do anything that was unplanned. There are various approaches that are made available.

In this paper, we will be aiming our attention at black-box testing and white-box testing. Black box testing which is also known as dynamic testing in the application includes testing on the behavior and functionalities of the application [25]. It uses a security testing tool to simulate an attacker and from there, the tool will identify the attack surface of the application [26]. Thus, the internal structure of the application will be excluded from black-box testing. An alternative way of doing black box testing other than using the attacker by simulation is by entering the inputs by the tester himself and obtaining the corresponding outputs. Hence, with this, bugs can be identified and modification of code helps in improving the security of the application in whole.

Another approach is white-box testing. White-box security testing which is also known as static testing includes testing on the internal structures of the application. The information mentioned will be extracted from the way the application is designed, design documentation and the source code of the application [24]. In white-box testing, internal security holes, poorly structured paths in coding, the flow of inputs, expected output, conditional loops and testing of each statement are involved. During the testing of predefined inputs against desired outputs, if there is an input that produces an unexpected output, a bug is encountered [27].

Some other security testing techniques include code-based testing, static analysis, dynamic analysis. Code review can be done manually or automated [24]. Manual code review is where an expert analyzes and checks the source code by going line by line to identify vulnerabilities. Therefore, a high-quality manual code review requires the expert to communicate with the software developers in order to get hold of the purpose and functionalities of the application. The output of the analysis will then be reported to the developers for further bugs fixing. Static analysis examines the source code without executing the program whereas dynamic analysis examines the source code when the program is executing [28]. Therefore, static analysis detects bugs at the implementation level while dynamic analysis detects errors during program runtime [9].

On the other hand, automated static application security testing (SAST) automatically analyzes and checks the source code. It will report potential vulnerabilities to the developer as soon as the source code is entered in the program. Unlike manual code review which requires line by line checking, SAST tools do not have limitations on the amount of program code, hence increasing the scalability of code review. Compared to the dynamic test approaches, SAST tools are producing a higher coverage and lower false negative rate, which is the reason why SAST is a more effective way of doing code reviews.

## 3 Data Collection

Data collection throughout this report is a crucial aspect as the collected result will impact the findings of the document. In principle, the data we collected is secondary data which the data were collected and tested previously by other investigators and researchers. By using the secondary data, it eases us to do research on this topic providing better results as we can utilize the data and save time on searching and investigating other related information. There are many sources of secondary data such as books, websites, published sources, newspapers, blogs and more [29]. However, in this report, we are going to utilize the research papers, e-books, websites and published sources.

As the topic of this paper is somewhat concerned by the developers who are going to manage the security. Thus, an abundance of associated information is available online. The primary source of the research database we used is Google Scholar. It is the most freely accessible research database. The Google Scholar is famous as all of the documents such as academic journals, books, conference papers and more are freely available. It is the strongest in technology fields nowadays as it collaborates with many technological partners, hence a lot of latest materials are being included [30].

Other than that, authors have utilized the ResearchGate database to acquire the applicable information about the topic. The ResearchGate database is commonly involved in the field of engineering and computer science which indicates to us to find information about the Secure Development Process for Developers [31]. Moreover, we also utilize the most broadly accepted format for research paper writing, the Institute of Electrical and Electronics Engineers (IEEE) as our source of data collection. The IEEE is well known and predominantly for the field of engineering and computing. It broadly consists of articles, scholarly references and conference papers. With their educational and professional cited publications, we are able to inquiry various kinds of related research papers on the IEEE research database. Last but not least, authors have also used the power of the internet resources to obtain additional information as our document analysis. The internet is the easiest and fastest way to acquire related information as it delivers a rich source of information about the secure development process for developers. However, the information we only approve about the topic was distributed by a credible and legitimate specialist.

In order to present a trustworthy research paper, there are many evaluation and countermeasures that need to be noted. When accessing the internet to obtain the information, we will check the date of the websites. As for reliable internet resources, the information will be updated frequently for a better understanding of their users. On the other hand, we will have a checking of the authors' information when accessing the article from the research database. By searching the author's experience and academic qualifications via the internet, we are able to identify whether the author is a legitimate professional in the field of the study. If the information we acquire from the authors is less qualified, the research paper will be not taken completely instead.

The default setting for each research database will look for the full text of the research paper. Nevertheless, when searching the keywords of "secure development process", "role of secure software development", "guideline for developers in software development" and more, the searching results will be sorted and generated within a short amount of time. Also, with the Boolean Operator in the search function, it is able to reduce or expand the results [32]. For example, if the keyword searching was "software development and design", the "AND" terms will return both searching terms that are included in the document. By using two terms "OR" or "AND" operators, the resulting will be more focused and productive. Correspondingly, the results will provide us with a comprehensive option to decide from. Moreover, a good rule of thumb is utilizing the research paper published within 5 years as it will be more reliable. Hence, sophisticated searching is used which the year range from 2015 to 2020 is set in order to narrow down the searching result.

## 4   Findings and Proposed Solution

*A.   Findings:*
In this section findings on what issues can surface if the techniques discussed previously are not applied to the whole process by the developers. This includes risk assessment, threat modeling, secure code review and security testing.

Risk assessment and threat modeling are carried out to limit the risk of software-based systems [33]. The threats and vulnerabilities are detected and identified in the early phases of SDLC, which are the planning and design phase. To build a feasible application, the potential threats of the targets must be studied as there are more and more new attack techniques being discovered, especially in smart but resource-constrained infrastructures like IoT [34][35]. Before analyzing the threats, the risks must be assessed. Following on, threat modeling minimizes the risk and its associated impacts. Risk assessment and threat modeling works together to assess the risks, analyze them and mitigate them. If risk assessment and threat modeling is not integrated in the SDLC, the end product may contain a lot of potential threats to be exploited by attackers.

Also, secure code review implementation is an essential step to ensuring the SDLC process is secure. This method is a necessary practice for secure software, however, it is only part of the solution [36]. Secure code review detects any possible vulnerabilities in the code before the launch of the product and this is to ensure that the code is free from these vulnerabilities and that it is also an optimized and efficient code. By applying this in the process of SDLC, the product has a competitive edge as it prevents users being at risk as the issue is managed during the development of the product hence, increasing reliability and reputation of said organization [37]. A case that often occurs from insecure code is evidently seen during release to the users. This method has been made a standard to some as it ensures the product is robust and that possible patches can be reduced as it can be prevented during the process which increases performance and work efficiency.

Lastly, with the developers not taking into account security testing into the phases will bring to a series of consequences. If a web application was released without security testing, it has the possibility of getting attacks such as SQL Injection, XSS and Brute Force Attacks. This case does not only apply to web applications but also software. Software is open to Buffer Overflow Attacks and Sniffing such as Brute Force, Malicious Code and Dictionary-Based Attacks [38]. Thus, integrating security testing into the development process of software is necessary as it ensures data security and system security, preventing data from leaking and unauthorized access.

From the analysis in previous sections it is observed that there is a lack of understanding and unawareness about how important these security implementations are to the phases; they can bring about many obstacles in an organization and its future projects, development processes and overall growth in the long run. That is why, by implementing the change from the beginning, the security of the SDLC will not only increase the organizations' efficiency but similarly it will increase its standing in their field as their reliability and assurance is more guaranteed as opposed to the products and software done by just implementation of a process without proper security guidelines.

*B.  Proposed Solution:*

The secured development process is an integral part in order to build a high-quality project. Nevertheless, the increasing complexity of the systems has created a few challenges for the developer to meet the high demand of the clients. Before the development of the secure software project, the developer should have security metrics. From the quote of "If you cannot measure it, you cannot improve it.", the developer needs to examine the security metrics in order to make the project efficient. Nonetheless, quantity is not equal to quality, it is not compulsory to measure everything. Instead, make it simple [39].

In implementing a secure software development, the developer might utilize the open-source elements such as libraries, frameworks or base code that are available for modification. As a consequence of being prebuilt, it can help the developer to reduce the development time. However, it would expose a vulnerability code risk and unwanted dependencies. Typically, a lot of open sources are utilized in a "black box" way. The developer only knows the appearance of the open-source elements instead of the operation. In order to resolve this challenge, the developer should only use the open-source elements in a situation that is unavoidable so that it can ensure the elements are not tampered by the attackers. The developer can also use the Software Composition Analysis (SCA) to detect the open-source code, third party application elements with the aim of security risk management, tracking vulnerabilities component and remediation automatically [40].

Besides that, the programming language used can become a challenge in secure software development as the language used by the developer will have its own unique vulnerabilities. By resolving this problem, the developer can make use of the APIs to filter the user interaction and hence reduce the opportunities to attackers [41]. Moreover, the buffer overflow in code is often neglected by the developer as they must save time and meet the deadlines of the project. This can result in the system crash if an attacker launches a buffer overflow attack [42]. Therefore, the developer should use languages that provide built-in protection. They can also utilize the Address Space Randomization (ASLR) to randomize the memory locations and ultimately make it difficult for the attacker to locate them [43]. The Data Execution Prevention (DEP) can be applied as well because it will mark the certain areas as non-executable and hence, the attacker will not be able to inject the code in that area.

As mentioned previously, security testing is vitally important to implement in each of the phases as it will help the developer who is unfamiliar with the security knowledge to remedy the code. There are a few software testing tools that can be utilized in different categories of possible issues such as Dynamic Security Testing (DAST), Static Analysis Security Testing (SAST) and Fuzz Testing [44]. The Fuzz Testing, one of the black box testing is used to identify the errors and vulnerabilities of the code such as Denial of Service (DoS), Cross-Site Scripting (XSS) and SQL injection. It will replicate the attack patterns by the attackers where they input the random data to test the code **[45]**. However, there should be more robust approaches developed to test the complex systems for more severe attacks like ransomware [46], and attacks targeted at systems involving resource-constrained devices [47].

## 5   Conclusion

In this age and era of technological advancement, the production of applications and software are ever abundant and organizations that are going with that tendency sometimes overlook the actions taken in achieving that goal especially in terms of security which in turn may lead to more concerns surfacing after the release of that product. Therefore, from our research, we have discovered the reasons behind the inconsistencies found within the organizations due to their view and approach in the SDLC process and at the same time presented its counter, which is the implementation of a few techniques and methods that can improve and create a more secure SDLC process for developers. In this paper, we also delved deeper into the findings and benefits of these implementations. Not only that, but the research also included the possible issues and challenges that can result from overlooking this aspect as well as its solutions. In conclusion, we have discussed and agreed that with the proper understanding and application of certain methods, a more efficient and cost effective process of the project can be achieved and at the same time the reliability of said software can be achieved and maintained. Lastly, a developer's role in the process of developing a product is crucial and it is best that these individuals get proper guidance and awareness in regards to how their knowledge, application and awareness can directly affect the outcome of the software especially during the SDLC process.